\documentclass[12pt]{article}
\usepackage{color,epsfig}
\usepackage{graphicx}
\usepackage[bf]{subfigure}

\definecolor{violet}{rgb}{0.4,0,0.6}
\definecolor{vert}{rgb}{0,0.6,0.2}
\definecolor{grun}{rgb}{0,0.8,0.0}
\definecolor{navy}{rgb}{0.0,0.0,0.6}
\definecolor{brun}{rgb}{0.4,0.0,0.0}

\def\spose#1{\hbox to 0pt{#1\hss}}\def\lta{\mathrel{\spose{\lower 3pt\hbox
{$\mathchar"218$}}\raise 2.0pt\hbox{$\mathchar"13C$}}}  \def\gta{\mathrel
{\spose{\lower 3pt\hbox{$\mathchar"218$}}\raise 2.0pt\hbox{$\mathchar"13E$}}}
\def\Libra{{\color{navy}{\spose {\lower 2.5pt\hbox{${^=}$}}{\cal L}}}}
\def\Euro{{\color{navy}{\spose {\lower 2.5pt\hbox{${^=}$}}{\bf C}}}}

\def\be{\begin{equation} } \def\fe{\end{equation}}

\def\bfa{ {\color{navy}{\bf a}} }\def\bfb{ {\color{navy}{\bf b}} }
\def\rPsi{ {\color{red} \Psi} }
\def\vaphi{ {\color{violet} \varphi} }

\def\vAA{ {\color{violet} A} }
\def\vPP{ {\color{violet} P} }
\def\vFF{ {\color{violet} F} }

\def\rkap{ {\color{red} \kappa} }
\def\rPhi{ {\color{red} \psi} }
\def\calL{ {\color{red} {\cal L}} }

\def\ee{ {\color{brun} e} }
\def\alfa{ {\color{brun} \alpha} }
\def\mm{ {\color{brun} m} }
\def\delth{ {\color{brun} \delta} }
\def\elth{ {\color{brun} \ell} }
\def\reta{ {\color{brun} \eta} }
\def\rhh{ {\color{red} h} }
\def\rff{ {\color{red} f} }

\def\varad{{\color{black} x}}
\def\quotient { {\color{brun} q }}

\def\lamb{ {\color{navy} \lambda} }
\def\gamb{ {\color{navy} \gamma} }
\def\gammatwo{ \gamb_{_{\bf 2}} }
\def\gammathree{ \gamb_{_{\bf 3}} }
\def\consthree{ {\color{navy} c}_{_{\bf 3}} }
\def\constwo{ {\color{navy} c}_{_{\bf 2}} }
\def\consone{ {\color{navy} c}_{_{\bf 1}} }
\def\conszero{ {\color{navy} c}_{_{\bf 0}} }

\def\calV{ {\color{red} {\cal V}} }

\def\calE{ {\color{red} {\cal E}} }

\def\UU{ {\color{red} U} }
\def\LL{ {\color{red} L} }

\def\vom{ {\color{violet} \omega} }
\def\vok{ {\color{violet} k} }

\def\vww{ {\color{violet} w} }

\begin{document}

\centerline {\color{red} {\Large The logarithmic equation of state 
for superconducting cosmic strings } }
\vskip 1.2 cm
\centerline{ {\bf \color{violet} Betti Hartmann 
\footnote{E-mail:b.hartmann@jacobs-university.de} }}
\vskip 0.3 cm
\centerline{\color{vert} School of Engineering and Science, 
Jacobs University Bremen, 28759 Bremen, Germany}
\vskip 0.6 cm
\centerline{ {\bf \color{violet} Brandon Carter 
\footnote{E-mail:brandon.carter@obspm.fr} }}
\vskip 0.3 cm
\centerline{\color{vert} LUTh, Observatoire de Paris, Meudon, France}
\vskip 0.6 cm
\centerline{1st November 2008}
\vskip 1.2 cm

\begin{center}
\noindent
{\bf Abstract} 
\smallskip
\end{center}

This investigation follows up the suggestion that the equation of state for
superconducting cosmic strings provided by Witten's prototype biscalar 
field model can be well represented by an effective Lagrangian of simple
logarithmic form depending on only 3 independent parameters. The numerical
work described here confirms the validity of this approximation, and 
initiates the evaluation of the 3 required parameters, as functions of the 
masses and other parameters specifying the underlying U(1) $\times$ U(1) 
scalar field model in the limit for which the relevant gauge coupling 
constants are small.In this limit, subject to calibration of the relevant 
length and mass scales, the scalar field model is characterised by just 3 
dimensionless ratios which (in order to provide conducting strings) must be 
subject to three inequalities (of which two have obvious analytic 
expressions). It is found here that when all three of these inequalities
are satisfied by a reasonably large margin, there is a simple empirical 
formula that can be used to provide a fairly accurate prescription for the 
algebraic dependence on these 3 dimensionless ratios of the 3 parameters 
required for the logarithmic equation of state.

\vfill\eject

\noindent
{\bf 1. Introduction}
\medskip

This work is a contribution to a program whose purpose is to provide 
appropriate equations of state -- expressible as prescriptions for an 
appropriate Lagrangian action density on the world sheet -- for 
conducting string models of the kind needed for description of 
Witten's bosonic superconducting string mechanism \cite{Witten85}, of 
which a potentially important consequence \cite{ShellardDavis89} is 
the formation of vortons. 

As a generalisation of the Nambu Goto prescription, which is simply a
constant, Witten originally suggested \cite{Witten85} that (in the 
absence of electromagnetic background fields) as a function of the 
quadratic norm, $\vww=\vaphi_{i}\,\vaphi^{i}$, of the surface gradient 
with components $\vaphi_{i}=\vaphi_{,\, i}$ of the relevant scalar 
phase field $ \vaphi$, the action density on the 2-dimensional string 
worldsheet might be taken to have the linear form
{\be \LL=-\mm^2-\frac{_1}{^2}\rkap_{_0} \vww\, , \label{Wit85}\fe}
depending on just two constant parameters $\mm$ (interpretable as a 
Kibble mass scale) and $\rkap_0$. It is evident that such a 
simplification can not provide a description of the effect of  
saturation when the current is large. However a more serious weakness 
is that even for the weak current limit in which $\vww$ is very 
small such a linear expression can not provide a satisfactory 
description of the propagation of small longitudinal perturbations, 
whose velocity depends not just on the first but also on the second 
derivative of $\LL$ with respect to $\vww$. 

To provide an acceptable description of perturbations in the weak 
current limit, as well as of what has been found \cite{Peter91} to occur 
when the current is larger, it has been proposed that, rather than
the Lagrangian itself, it is the inverse of the derived quantity
{\be \rkap=-2\frac{{\rm d}\LL}{{\rm d}\vww} \label{kap}\fe}
that should be taken to be linear. This means that the Witten formula 
(\ref{Wit85}) should be replaced by an expression of the logarithmic form
{\be \LL=-\mm^2-\frac{_1}{^2}\mm_\star^{\,2}\, {\rm ln}\,
\{1+\delth_\star^{\,2}\,\vww\}\ \, ,\label{logarithmic}\fe}
which depends on three fixed parameters, of which one is the Kibble mass 
scale, $\mm$ (as before) while the others are a secondary mass scale 
$\mm_\star$ and a lengthscale $\delth_\star$, whose squared product is 
identifiable with the original Witten model parameter, $\rkap_{_0}$.
The latter is interpretable as the zero current value of the quantity 
(\ref{kap}) which for non zero values of $\vww$ will be given by
{\be \rkap=\frac{\rkap_{_0}}{1+\delth_\star^{\,2}\,\vww} \, ,
\hskip 1 cm \rkap_{_0}= \mm_\star^{\,2}\delth_\star^{\,2}\, .
\label{condi}\fe}

Models of this kind are a special subclass within the general category 
needed for studying properties of generic conducting string models, of 
which an important example is the analysis 
\cite{CarterMartin93,Martin94,MartinPeter95,CarterPeterGangui97}
of the stability of their vorton (closed loop) equilibrium states. Whether
the particular logarithmic form (\ref{logarithmic}) 
will be able to provide a good description for conducting string defects 
in realistic field models such as those of the standard electroweak theory 
and its various extensions \cite{DavisPerkins97,Volkov06} is a problem
that will be left for future work. The present article will be
concerned just with its application in the restricted
 class of toy U(1)$\times$ U(1) field models proposed in this
context by Witten. 

Our purpose here is thus to investigate the dependence of the three 
parameters $\mm$,  $\mm_\star^{\,2}$, $\delth_\star^{\,2}$, in the 
logarithmic string model on the various parameters needed in Witten's 
class \cite{Witten85} of toy field models. It was already recognised
in the earliest pioneering exploration of the relevant parameter 
space \cite{Babul88} that there will only be a restricted part 
of it for which the non-vanishing currents envisaged by Witten will 
actually be able to occur. Outside the parameter domain of this  
restricted Witten subclass, the ensuing string model will merely be of 
the Nambu Goto type, as obtained from the Witten formula (\ref{Wit85}) 
by setting $\rkap_{_0}=0$, or equivalently from  the logarithmic formula 
(\ref{logarithmic})  by setting $\mm_\star=0$.

One of the main advances in the present work will be to provide a much 
more accurate and complete description of the conditions that are 
necessary and sufficient for the characterisation of this Witten domain 
within the entire parameter space of the Witten class of toy scalar 
field models. Near the boundaries of this domain, the parameter dependence 
of the quantities characterising the conducting string model turns out to 
be rather sensitive. However better behaviour 
is obtained whenever the necessary inequalities are all satisfied with 
a reasonably large margin, and in such generic circumstances it is found 
that simple algebraic formulae can be provided for the specification 
with reasonable accuracy of the empirical string model parameters $\mm$,  
$\mm_\star^{\,2}$, $\delth_\star^{\,2}$ as functions over the Witten 
domain in the parameter space  of the underlying scalar field model.

\bigskip
\noindent
{\bf 2. Specification of the class of field models}
\medskip

The Witten class of superconducting string models, to which the present
work applies, is characterised by appropriate parameter restrictions within 
a more  complete category that has recently been set up for a different 
purpose by Saffin \cite{Saffin05}, whose concern was with another part of 
the relevant parameter space where it gives rise to strings that are 
non-conducting but that are instead characterised by not just one but 
two independent winding numbers, a feature that gives rise to the 
possibility of string junctions \cite{Copeland05,Copeland07}.

The complete U(1)$\times$ U(1) category \cite{Saffin05} consists of Abelian 
models involving a pair of complex scalar fields $\rPsi_\bfa$ with label 
$\bfa=1, 2$ that are coupled by a potential $\calV$ but subject to a 
Lagrangian of the (rationalised) form
{\be \calL=-\sum_\bfa\left(\frac{_1}{^2}({\cal D}_{\!\mu}
\rPsi_\bfa){\cal D}^{\mu}\overline\rPsi_\bfa +\frac{1}{4}\vFF^\bfa_{\,\mu\nu}
\vFF^{\bfa\mu\nu}\right) -\calV\, ,\fe}
in which the kinetic part is entirely decoupled, meaning that each of the 
scalar fields $\rPsi_\bfa$ is subject to the gauge action only of its own 
corresponding Abelian connection $\vAA^\bfa_{\!\mu}$ with corresponding field 
$\vFF^\bfa_{\!\mu\nu}=2\nabla_{\![\mu}\vAA^\bfa_{\nu]}$ according to the 
simple specification whereby each has its own charge coupling constant 
$e_\bfa$ in terms of which ${\cal D}_{\!\mu}\rPsi_\bfa=(\nabla_{\!\mu} 
\rPsi_\bfa- i \ee_\bfa \vAA^\bfa_{\!\mu})\rPsi_\bfa$. Each complex scalar is 
expressible in the  usual manner, $\rPsi_\bfa=\rPhi_\bfa\, {\rm exp}
\{i\vaphi_\bfa\} $ in terms of a real phase angle $\vaphi_\bfa$ and a real 
amplitude $\rPhi_\bfa$. Only the amplitudes are involved in the potential, 
which is taken to be given in terms of a pair of mass scales 
$\tilde\reta_\bfa$ by an expression of the quartic form 
{\be \calV =\frac{1}{4}\sum_{\bfa\bfb} \lamb^{\bfa\bfb}(\rPhi_\bfa^2
-\tilde\reta_\bfa^2) (\rPhi_\bfb^2-\tilde\reta_\bfb^2)+ \calV_{_0}\fe}
in which the  $\lamb^{\bfa\bfb}$ are the three independent components of a 
symmetric matrix that is restricted by the condition that $\calV$ must be 
bounded below, and $\calV_{_0}$ is a dynamically redundant constant included 
for the purpose of adjusting the lower bound of $\calV$ to be zero. This 
condition evidently requires that the diagonal coefficients should both be 
positive $\lamb^{_{\bf 11}}\geq 0\, ,\ \ \lamb^{_{\bf 22}}\geq 0$. It can be 
seen that a further necessary and sufficient condition for the potential to 
be bounded below is that the off diagonal component should be subject to the 
lower bound $\lamb^{_{\bf 12}}\geq -\sqrt{\lamb^{_{\bf 11}}
\lamb^{_{\bf 22}}}$. Saffin's concern \cite{Saffin05} was with the class 
characterised by positivity of the determinant $\lamb^{_{\bf 11}}
\lamb^{_{\bf 22}}-(\lamb^{_{\bf 12}})^2$ itself, for which (with 
$\calV_{_0}=0$) the vacuum is characterised by simultaneously non-vanishing 
field values $\rPhi_{_{\bf 1}}=\tilde\reta_{_{\bf 1}}$ and $\rPhi_{_{\bf 2}}=
\tilde\reta_{_{\bf 2}}$.

It is the opposite alternative possibility with negative determinant,
as specified by the more restrictive lower bound 
{\be \lamb^{_{\bf 12}}>\sqrt{\lamb^{_{\bf 11}}\lamb^{_{\bf 22}}}
\label{detneg}\fe}
that characterises models of the Witten class with which we are concerned 
here. In this case the potential has minima where one or other of the fields 
vanish, specifically -- setting $ \reta_{_{\bf 1}}^{\,2}
=\tilde\reta_{_{\bf 1}}^{\,2}+\lamb^{_{\bf 12}}\tilde\reta_{_{\bf
    2}}^{\,2}/\lamb^{_{\bf 11}}$  and $\reta_{_{\bf 2}}=
\tilde\reta_{_{\bf 2}}^{\,2}+ \lamb^{_{\bf 12}}\tilde
\reta_{_{\bf 1}}^{\,2}/\lamb^{_{\bf 22}}$ --
where $\rPhi_{_{\bf 1}}=0$ with $\rPhi_{_{\bf 2}}=
\reta_{_{\bf 2}}$, and where  $\rPhi_{_{\bf 2}}=0$ with $\rPhi_{_{\bf 1}}
=\reta_{_{\bf 1}}$. 
The vacuum state itself (excluding the degenerate case for which the 
minima are equal, which gives a model with domain walls) will be the 
one for which $\calV$ is lowest. Without loss of generality  this can 
be taken to be the latter, where $\rPhi_{_{\bf 2}}=0$,  by choosing the 
labelling so that the new mass scales satisfy
{\be \lamb^{_{\bf 11}}\,\reta_{_{\bf 1}}^{\,4}
> \lamb^{_{\bf 22}}\, \reta_{_{\bf 2}}^{\,4}\, ,\label{ineq}\fe}
 rather than the other way round. With his convention we shall be
left with
{\be \calV =\frac{\lamb^{_{\bf 11}}}{4}\big(\rPhi_{_{\bf 1}}^{\,2}
-\reta_{_{\bf 1}}^{\,2}\Big)^2 + \frac{\lamb^{_{\bf 12}}}{2}
\rPhi_{_{\bf 1}}^{\,2}\rPhi_{_{\bf 2}}^{\,2} + \frac{\lamb^{_{\bf 22}}}{4}
\rPhi_{_{\bf 2}}^{\,2}(\rPhi_{_{\bf 2}}^{\,2}-2\reta_{_{\bf 2}}^{\, 2})
\, . \label{potlform}\fe}
after suitable adjustment of the additive constant 
The effective masses, $\mm_{_{\bf 1}}$ and $\mm_{_{\bf 2}}$ say, of the 
primary (Higgs type) and secondary (carrier) fields (namely 
$\rPsi_{_{\bf 1}}$  and $\rPsi_{_{\bf 2}}$) will then be given respectively by
the formulae
{\be \mm_{_{\bf 1}}^{\,2}=\lamb^{_{\bf 11}}\reta_{_{\bf 1}}^{\,2}\, ,
\hskip 1 cm\mm_{_{\bf 2}}^{\,2}=\lamb^{_{\bf 12}}\,\reta_{_{\bf 1}}^{\,2}-
\lamb^{_{\bf 22}}\reta_{_{\bf 2}}^{\,2} \, ,\label{masses}\fe}
the requisite positivity of the latter being guaranteed as an automatic 
consequence of the inequalities (\ref{detneg}) and (\ref{ineq}).

The lengthscale
{\be \delth=\frac{1}{\mm_{_{\bf 1}}}=\frac{1}{\reta_{_{\bf 1}}
\sqrt{\lamb^{_{\bf 11}}}} \label{lengthscale} \fe}
associated with the primary field will roughly characterise the 
effective core radius of the ensuing string type vacuum defects. In order 
for such  string defects to be strictly local (with finite world sheet 
energy density) we now follow Peter \cite{Peter91} in restricting our 
attention cases in which there is no coupling to any external electromagnetic 
field by setting $\ee_{_{\bf 2}}=0$. Far outside a string defect the field 
configuration will then converge towards the vacuum with an exponential 
cut off lengthscale $\elth$ (of the kind known in the context of ordinary 
metallic superconductivity as the London length) given in  terms of the 
coupling constant $\ee_{_{\bf 1}}$ of the other (non-electromagnetic) 
gauge field by the formula
{\be \elth^2=\frac{1}{e_{_{\bf 1}}^{\, 2}\reta_{_{\bf 1}}^{\,2}}\, .\fe} 
The well known Bogomolny limit condition -- in the case for which the
carrier field is absent -- is simply that these lengthscales be equal
$\ell=\delta$, which occurs when the coupling has the critical value
$\ee_{_{\bf 1}}^{\, 2}= 2\lamb^{_{\bf 11}}$. However our concern here
will be with the weak coupling limit within the type II range 
characterised by the condition $\elth>\delth$. It is convenient to write
{\be \alfa=\sqrt\frac{\ee_{_{\bf 1}}^{\, 2}}{\lamb^{_{\bf 11}}}
\, ,\fe} so as to define a
positive dimensionless charge coupling constant $\alpha$ in terms of which
this type II range is characterised by the condition $\alfa<\sqrt{2}$.
In addition to this dimensionless parameter $\alpha$, the system will
be fully characterised qualitative, meaning modulo rescaling, by three
other dimensionless parameters, which can be conveniently taken to
be the mass ratio
{\be \quotient=\frac{\reta_{_{\bf 2}}}{\reta_{_{\bf 1}}}\, ,\fe}
and the pair of ratios
{\be \gammatwo=\frac{\lamb^{_{\bf 22}}}{\lamb^{_{\bf 11}}}\, ,\hskip 1 cm
 \gammathree=\frac{\lamb^{_{\bf 12}}}{\lamb^{_{\bf 11}}}\, .\fe}
In terms of these, the necessary (but as we shall see, not quite 
sufficient) inequalities (\ref{detneg}) and (\ref{ineq}) characterising the 
Witten subclass can be written conjointly as the condition
{\be \quotient^4\gammatwo^{\,2}<\gammatwo<\gammathree^{\,2} 
\, .\label{Witcons}\fe}

\bigskip
\noindent
{\bf 3. Cylindrical vortex configurations}
\medskip

The string model for a vortex defect of the vacuum is based
on the assumption that a small segment thereof can be
represented by a stationary cylindrically symmetric
configuration in which, in terms of constants 
$\vom$, $\vok$, $n$, of which the latter is an integer winding number, 
the fields  will be specified with respect
to cylindrical coordinates $\{t,r,\theta, z\}$ by the ansatz
{\be \ee_{_{\bf 1}}\vAA_\mu\, {\rm d} x^\mu \! =\!(n-\!\vPP)\, {\rm d}\theta 
\, ,\hskip 0.1 cm \rPsi_{_{\bf 1}} \! = \reta_{_{\bf 1}} \rhh \ {\rm exp}
\{i n \, \theta\}\, ,\hskip 0.1 cm \rPsi_{_{\bf 2}} \! = \reta_{_{\bf 1}}
 \rff \, {\rm exp}\{i \vok z- i \vom\, t\}\, , \fe}
in which the quantities $\vPP$, $\rhh$, $\rff$  depend only
on the radial coordinate $r$, and are subject to the boundary conditions
{\be\vPP \rightarrow n\, ,\hskip 0.6 cm  \rhh \rightarrow 1\, ,\hskip 
0.6 cm \rff \rightarrow 0\, ,\hskip 0.6 cm   {\rm as} \ \
r\rightarrow\infty\, ,\fe}
{\be \vPP \rightarrow 0\, ,\hskip 0.6 cm \rhh \rightarrow 0\, ,
\hskip 0.6 cm \frac{ {\rm d}\rff }{ {\rm d} r}\rightarrow 0\, ,
\hskip 0.6 cm    {\rm as} \ \ r\rightarrow\infty\, .\fe}
 The corresponding macroscopic string
description, of the kind discussed in the introduction, will be
obtained by taking
{\be \LL= 2\pi\!\int\! \calL\, r\,{\rm d} r\, ,\hskip 1 cm
\rkap  = 2\pi\!\int\! \rPhi_{_{\bf 2}}^{\,2}
\, r\,{\rm d} r\, ,\hskip 1 cm \vww=\vok^2-\vom^2\, ,\label{integrals}\fe}
subject to the   field equations. Using a prime for differentiation with 
respect to the dimensionless radial coordinate defined in terms of the 
lengthscale (\ref{lengthscale}) as 
{\be \varad=\mm_{_{\bf 1}}\, r\, ,\fe}
 these field equations will be given by
{\be \left(\frac{\vPP^{\,\prime}}{\varad}\right)^\prime=\alfa^2 
\frac{\vPP\rhh^2}{\varad}\, ,\fe}
{\be \frac{1}{\varad}(\varad\,\rhh^\prime)^\prime=\frac{\vPP^2\rhh}{\varad^2}
+\rhh(\rhh^2-1)+\gammathree\,\rff^2\rhh\, ,\fe}
{\be \frac{1}{\varad}(\varad\,\rff^\prime)^\prime=\frac{\vww}
{\mm_{_{\bf 1}}^{\,2}}\,\rff +\gamma_2\, \rff(\rff^2-\quotient^2)
+\gammathree\,\rff\rhh^2\, .\fe}

\bigskip
\noindent
{\bf 4. Procedure for empirical matching of the equation of state}
\medskip

Numerical application of the foregoing procedure can in principle
provide an exact equation of state specifying the dependence of
$\LL$ on $\vww$. The suggestion \cite{CarterPeter95} that
the simple logarithmic formula (\ref{logarithmic}) can provide
a good approximation to the exact result was motivated by
the earlier analysis of Peter \cite{Peter91} which shows that
a singularity will occur when the energy associated with the
current reaches the threshold for creation of particles of the carrier
field with the mass $\mm_{_{\bf 2}}$ given by (\ref{masses}),
which can be seen to occur when $\vww=- \mm_{_{\bf 2}}^{\,2} \, .$
On this basis, the appropriate value for the parameter $\delth_\star$ in 
the equation of state can be evaluated in advance simply as 
{\be \delth_\star=\frac{1}{\mm_{_{\bf 2}}}\, .\label{delstar}\fe}

To complete the specification of the logarithmic formula 
 (\ref{logarithmic}) it remains only to obtain the mass parameters
$\mm$ and $\mm_\star$ which will both be obtainable from knowledge
just of the zero current limit, by evaluation of the corresponding
values $\LL_{_0}$ and $\rkap_{_0}$ as given by the integral formulae 
(\ref{integrals}) for $\LL$ and $\rkap$ when $\vww=0$,
from which one will obtain
{\be \mm^2=-\LL_{_0}\, ,\hskip 1 cm 
\mm_\star^{\,2}=\mm_{_{\rm 2}}^{\,2}\rkap_{_0}\, .\fe}   
This case $\vww=0$  includes the strictly static 
configuration characterised by $\vom=0$ as well as $\vok=0$, for which
the field equations will be obtainable just by minimisation of the 
corresponding energy integral. In view of the staticity the latter
will be equal in magnitude but opposite in sign to the Lagrangian, which 
takes the form 
{\be \LL_{_0}= -\reta_{_{\rm 1}}^{\,2}\,\tilde\UU\, ,\hskip 1 cm
\tilde\UU=2\pi\!\int\! \tilde\calE\,\varad\,{\rm d}\varad \, , \fe}
in which the dimensionless rescaled energy density is given by
{\be 2\,\tilde\calE=\frac{\vPP^{\,\prime\,2}}{\alfa^2\varad^2}+
\frac{\vPP^{\,2}\rhh^2}{\varad^2}+\rhh^{\prime\, 2}+\frac{(\rhh^2-1)^2}{2}+
\gammathree\rhh^2\rff^{\,2}+\rff^{\,\prime\, 2}+\gammatwo
\frac{\rff^{\,4}}{2}-\gammatwo\,\quotient^2\rff^{\,2}\label{En}\, .\fe}

Subject to the minimisation of the total energy integral $\tilde\UU$, 
the corresponding expression for the required condensate integral will be
{\be \rkap_{_0}=2\pi \frac{\reta_{_{\rm 1}}^{\, 2}}{\mm_{_{\rm 1}}^{\,2}}
\,\int\!\rff^{\, 2}\varad\,{\rm d}\varad \, .\label{knaught}\fe}
We thus end up with the integral prescriptions
{\be \frac{\mm^2}{\reta_{_{\rm 1}}^{\, 2}}=\tilde \UU\, ,\hskip 1 cm
\frac{\mm_\star^{\,2}}{\reta_{_{\rm 1}}^{\, 2}}=(\gammathree-\gammatwo
\quotient^2)2\pi\!\int \!\rff^{\,2}\varad\,{\rm d}\varad \, .\fe}

\bigskip
\noindent
{\bf 5. The third restriction}
\medskip

It is to be noticed that all the terms in (\ref{En}) are positive except the 
last, which is homogeneously quadratic in $\rff$, and that it is only where 
the magnitude  of this last term exceeds that of the other terms quadratic 
in $\rff$ that local minimisation will leave a non vanishing value for this 
condensate amplitude $\rff$. In particular this will occur only where 
$\gammathree\, \rhh^2<\gammatwo\,\quotient^2\, ,$ which for winding
number $n=1$ means roughly where $\gammathree\,\varad^2\lta
\gammatwo\,\quotient^2\, ,$ since it is to be expected that the order 
of magnitude of the primary field will be given roughly by $\rhh\simeq 1$ 
for $\varad\gta 1$ and $\rhh\approx \varad$ for  $\varad\lta 1 \, .$ The 
implication that non-vanishing values of $\rff$ must be roughly confined
within the thin tube characterised by $\varad^2\lta q^2\gammatwo/\gammathree$ 
entails that the squared fractional gradient of $\rff$ in the tube should on 
average satisfy $(\rff^{\,\prime}/\rff)^{2}\gta \gammathree/
(\gammatwo \quotient^2)\, .$ However this conflicts with
the requirement that the term $\rff^{\,\prime\,2}$ in (\ref{En}) should also 
be dominated by the final term, $-\gammatwo\quotient^2\rff^2\, ,$ unless
we have
{\be \gammathree\lta \gammatwo^{\,2}\quotient^4\, , \label{xtra}\fe}
which, in view of (\ref{Witcons}) entails that we must also have
$ \gammathree\gta 1 \, .$

\begin{figure}
\centering
\epsfig{figure=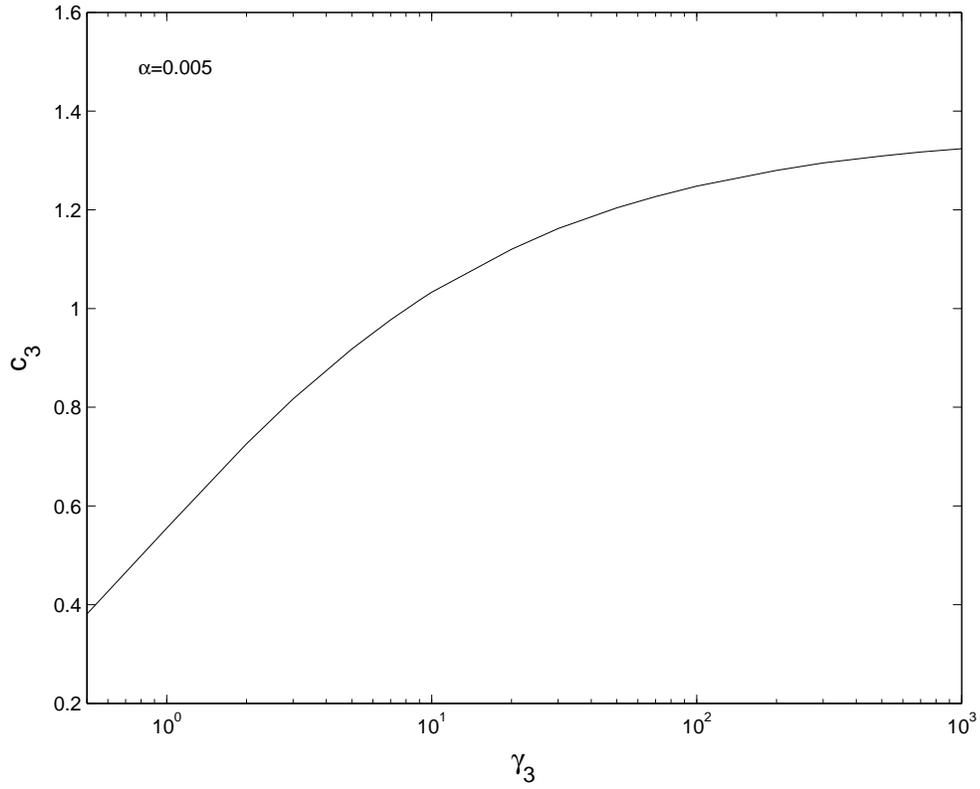,width=15.0 cm}
\caption{\label{fig_c3} The value of  $c_{\bf 3}$ (defined as the
lower bound for $\quotient^4\gammatwo^{\,2}/\gammathree$) is plotted 
(for weak gauge coupling, $\alpha=0.005$) as a function of 
$\gammathree$ over the allowed range of the latter, which
must always be greater than about 0.3.}
\end{figure}

The implication of this is that, in order for the secondary
field to provide a non vanishing condensate, as measured by
the sectional integral $ \rkap_{_0}$ given by (\ref{knaught}),
and thus as a prerequisite for string conductivity,
the two  inequalities in (\ref{Witcons}) are not sufficient. To fully
characterise the required Witten subclass, they must be
be supplemented by a third parameter restriction that will evidently
be given roughly by (\ref{xtra}) and that has been obtained in a
more precise form by numerical work.

This third restriction may be expressed in conjunction with
the other two of the form
{\be \consthree\gammathree<\gammatwo^{\,2}\quotient^4
<\gammatwo<\gammathree^{\,2} \, .\label{combi}\fe}
in which $\consthree$ is a dimensionless quantity of order unity.
Whereas we had expected that it might also depend weakly on 
$\gammatwo$ or on $q^2$ our numerical computations have shown that
(in the weak gauge coupling limit considered here)
 the lower bound  $\consthree$ can
be specified with high accuracy as a function only of $\gammathree$. 
The result obtained for the weak dependence of $\consthree$ on
$\gammathree$ is plotted in Fig.\ref{fig_c3}, from which it can be
seen that for large values of $\gammathree$ it increases towards a value 
given roughly by $\consthree \simeq  1.4$.

\bigskip
\noindent
{\bf 6. Formula for the generic (non-marginal) case.}
\medskip

Provided the conditions (\ref{xtra}) are satisfied not just marginally but by
a wide margin, the gradient term $\rff^{\,\prime\,2}$ in (\ref{En})
will be negligible so the contribution from $\rff$ will be effectively
algebraic and thus minimisable locally, providing the approximation
$\rff^{\,2}\simeq \quotient^2-\rhh^2 \gammathree/\gammatwo$ in the tube
where this is positive, and $\rff^2 \simeq 0$
outside. Again on the supposition that we shall have
$\rhh\approx \varad$ inside the tube (whose radius will therefor
be given by $x\approx \quotient\sqrt{\gammatwo/\gammathree}$)
this provides the corresponding rough order of magnitude estimate
{\be \gammathree\ll \gammatwo^{\,2}\quotient^4\hskip 1 cm \Rightarrow
\hskip 1 cm \int \!\rff^{\,2}\varad\,{\rm d}\varad \approx 
\frac{\quotient^4\,\gammatwo}{4\,\gammathree}\fe}
for the integral needed in the formula (\ref{knaught}) for the 
evaluation of  $ \rkap_{_0}$.

\begin{figure}[htbp]
\centering
{
    \includegraphics[width=11cm, height=6cm]{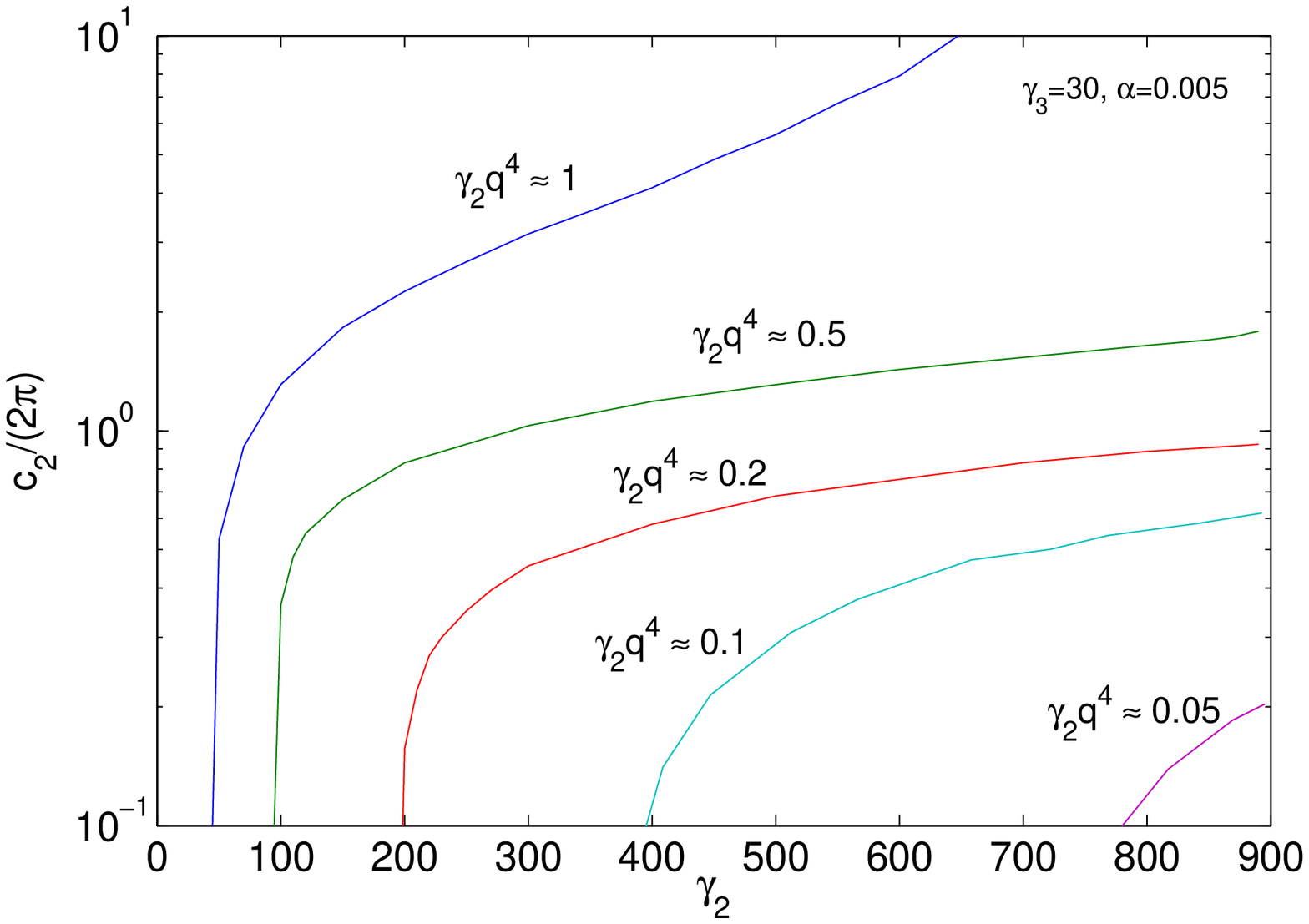} 
  }
{
    \includegraphics[width=11cm,height=6cm]{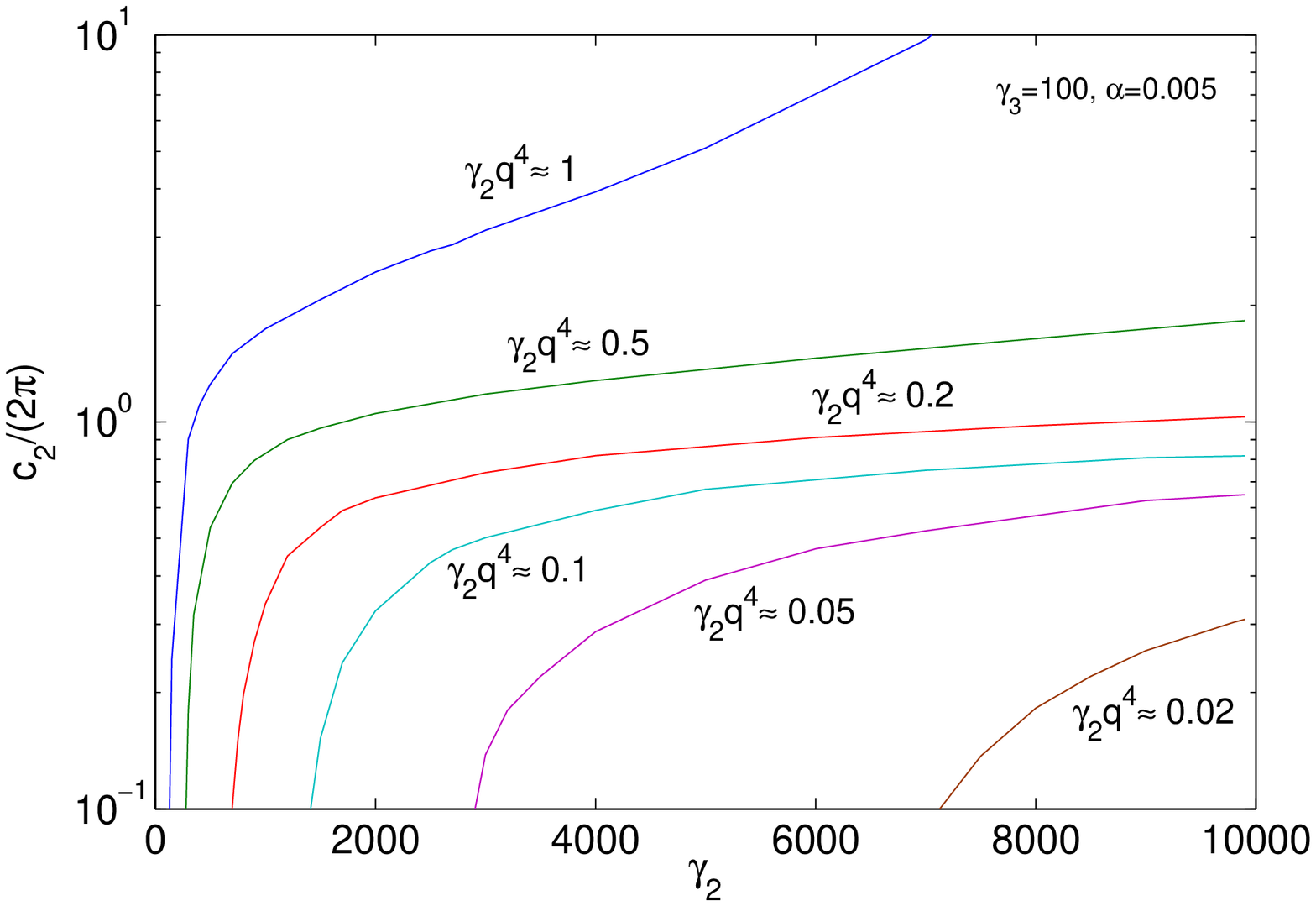} 
  }
 {
    \includegraphics[width=11cm,height=6cm]{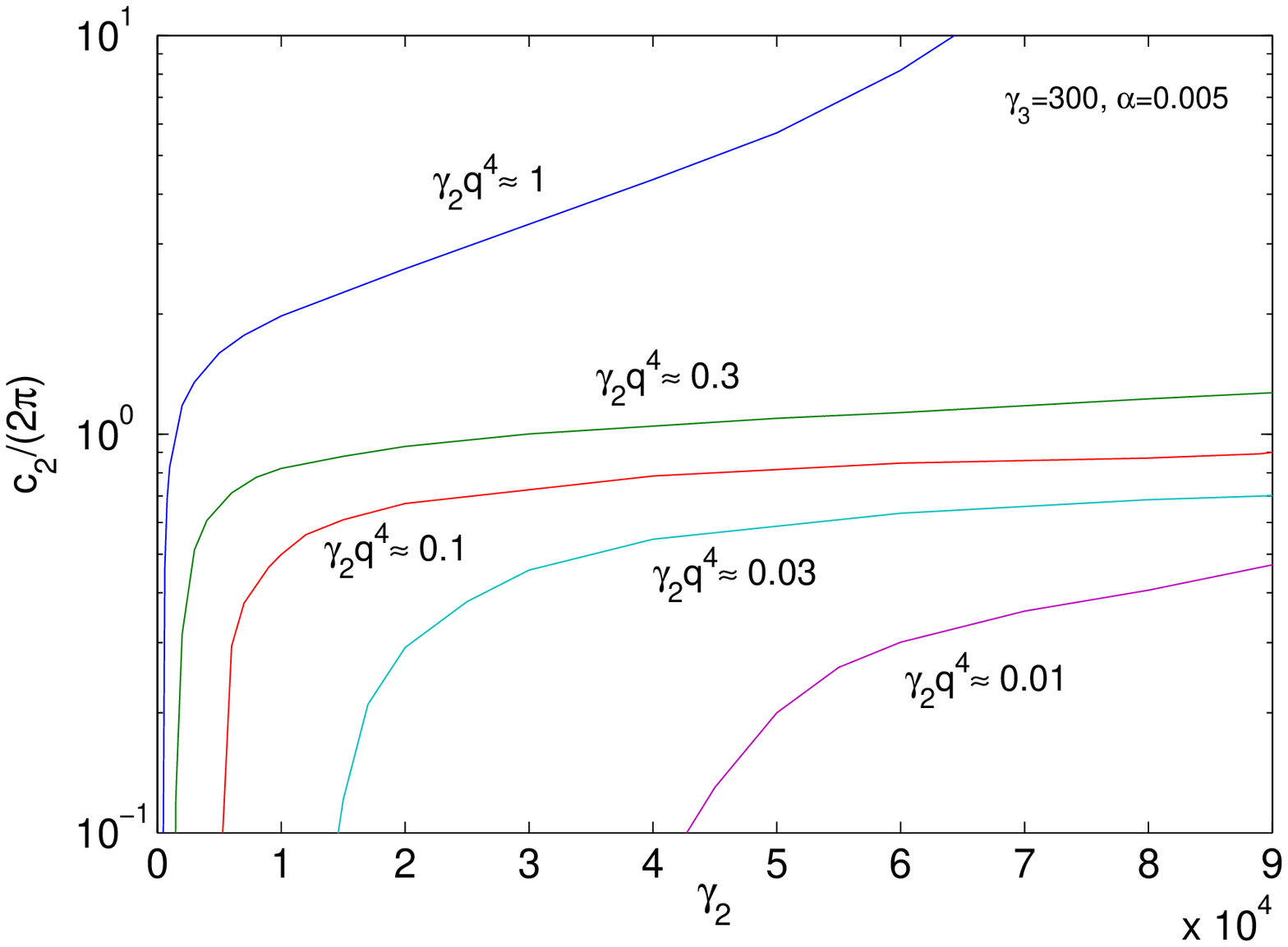} 
  }
\caption{The value of $\constwo/(2\pi)$ is plotted against $\gamma_2$ on 
chosen contours of constant  $\gammatwo \quotient^4$ (with
$\alfa=0.005$) for the successive values 30, 100, 300 of $\gammathree$.
 }
  \label{fig0}
\end{figure}

The implication is that this quantity
will be given by a formula of the form
  {\be \rkap_{_0}=\constwo \frac{\gammatwo\reta_{_{\bf 1}}^{\, 2} \quotient^4}
 {\gammathree\mm_{_{\rm 1}}^{\,2}} = \constwo\frac{
\lamb^{_{\bf 22}} \quotient^2 }{  \lamb^{_{\bf 11}}
\lamb^{_{\bf 12}} } \, ,
\label{kapform}\fe}
in which $\constwo$ is a coefficient that will be non-zero only  when the
 condition (\ref{xtra}) is satisfied, and that will tend,
when (\ref{xtra}) is satisfied by a wide margin, towards a roughly
constant order of unity value, $ \constwo\approx 1 \ \  .\label{cthree} $

In order to verify this, and to obtain a more precise estimate for the
coefficient in (\ref{kapform}) it is convenient to use this equation
as a formal definition for the parameter $c_{\bf 2}$,
in terms of the quantity  $\rkap_{_0}$ given by (\ref{knaught})
 as the integral of $\rff^2$ for a chiral solution. On the basis of this
definition, we have evaluated $\constwo$ as a function of
given $\gammatwo$, and $\quotient^2$ for various fixed values of 
$\gammathree$ taken well within the allowed range, that is for
$\gammathree\gg 1$. The results for the particular values
$\gammathree=30$, $\gammathree=100$, and $\gammathree=300$ are
given in the three successive plots of  Figure \ref{fig0}, 
which show contours for fixed values of the quantity $\gammatwo\quotient^4$ 
at levels up to its marginal upper limit $\gammatwo\quotient^4=1$.
So long as  $\gammatwo\quotient^4$ is neither too close to
this upper limit nor too near its lower bound (which decreases as
$\gammathree$ increases) it can be seen that over a wide range of $\gammatwo$ 
the value of $\constwo$ is roughly constant with a value of the ratio
$\constwo/(2\pi)$ that is close to unity, so that a reasonable approximation
will be obtained by taking
{\be \constwo\simeq 2\pi \, .\fe}

With $\delth_\star$ given in advance by ($\ref{delstar}$), and 
with the parameter $\rkap_{_0}$ -- and hence also 
$\mm_\star^{\,2}=\rkap_{_0}/\delth_\star^{\,2}$ -- determined by the
coefficient $\constwo$, the only quantity
still needed for the complete specification of the 
logarithmic equation of state (\ref{logarithmic}) is the
value of the Kibble mass $\mm$ itself. For this, dimensional
considerations suggest,  that a reasonably good description will
be provided by an expression of the form
{\be  \mm^2=\consone\, \reta_{_{\rm 1}}^{\, 2}\, {\rm ln}\left\{
\frac{\conszero}\alfa \right\} \, , \label{logasy}\fe}
in which, when the coupling is weak, $\alfa^2\ll 1 $, 
 the coefficients will have roughly constant order of unity values.

\begin{figure}
\centering
\epsfig{figure=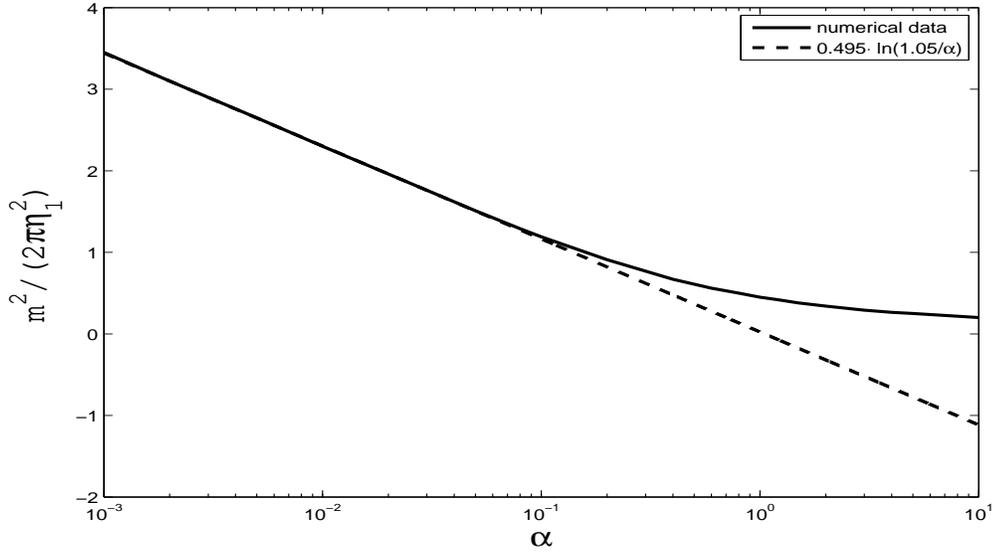,width=15.0 cm,height=8 cm}
\caption{\label{fig1} The value of  $\mm^2/(2\pi\reta_1^2)$ is plotted
against the coupling constant $\alfa$, using  a straight dashed line for the 
prediction of the equation (\ref{logasy}) and a solid curve for the numerical data,
which can be seen to fit the suggested form (36) very well for small
$\alfa$ but to deviate -- as expected -- for  $\alfa \gta 0.1$.
}
\end{figure}

As shown by Figure \ref{fig1}, his expectation seems to be confirmed by 
our numerical computations which have given the values
{\be \consone\simeq 3.11 \ \ , \ \  \conszero\simeq 1.05 \, .
\label{numres}\fe}.

\vfill\eject

\bigskip
\noindent
{\bf 7. Conclusions}
\medskip

In this study of superconducting strings in the weakly coupled 
limit ($\alfa\ll 1$) of the U(1)$\times$ U(1) field model of Witten, 
it has been found that
in order to provide such conducting strings, the parameters 
characterising the field model  must satisfy
the  necessary and sufficient conditions given by (\ref{combi}).
The applicability, as a reasonably good approximation, of 
the previously suggested logarithmic form (\ref{logarithmic})
for the effective Lagrangian characterising the equation of state has
been confirmed.  In terms of the parameters characterising the field
model by the specification of the relevant potential (\ref{potlform}), 
the required length scale  $\delth_\star$ will be given by (\ref{delstar})
and our numerical results (\ref{numres})
suggest that the relevant Kibble mass scale $\mm$ (which is all that
is needed in the non-conducting Nambu Goto limit) may be taken
with reasonable accuracy to be given, according to (\ref{logasy}),
by an ansatz of the easily memorable form
{\be \label{meq} \mm^2=\pi\, \reta_{_{\rm 1}}^{\, 2}\, {\rm ln}\left\{
\frac{1}{\alfa} \right\} \, ,\fe}
For the condensate integral $\rkap_{_0}$ needed to complete the 
specification of the equation of state by providing
the other required mass scale $\mm_\star$ using (\ref{condi}), the
formula (\ref{kapform}) provides a corresponding estimate
{\be \rkap_{_0}=2\pi \frac{
\lamb^{_{\bf 22}} \quotient^2 }{  \lamb^{_{\bf 11}}
\lamb^{_{\bf 12}} } \, ,\fe}
that will be valid as a reasonable approximation when the conditions
(\ref{combi}) are satisfied with a large margin, i.e. when
{\be \lamb^{_{\bf 11}}\lamb^{_{\bf 12}}\ll(\lamb^{_{\bf 22}}\quotient^2)^2
\ll  \lamb^{_{\bf 11}}\lamb^{_{\bf 22}}\ll (\lamb^{_{\bf 12}})^2 \, .\fe} 
\medskip

\noindent
{\bf Acknowledgments.}
\smallskip

 We wish to thank X. Martin and M. Volkov for illuminating conversations.
This work was supported by A.N.R prorgam NT05.1-42846. We also thank the ICTS visitors
program of Jacobs University Bremen for financial support.

\end{document}